\pdfoutput=1 \pdfsuppresswarningpagegroup=1 
\documentclass[
aps,
prb,
reprint, twocolumn,
groupedaddress,
superscriptaddress,
amsfonts,
letterpaper,
footinbib,
floatfix,
longbibliography
]{revtex4-2}

\RequirePackage{amsmath,amssymb,bm}
\RequirePackage{graphicx}
\RequirePackage[usenames,dvipsnames]{xcolor}
\usepackage{float}

\usepackage{color}

\usepackage{pdfpages}
\makeatletter\AtBeginDocument{\let\LS@rot\@undefined}\makeatother

\usepackage[T1]{fontenc}

\usepackage{enumitem}

\usepackage{verbatim}

\usepackage{hyperref}
\hypersetup{
    colorlinks=true,
    citecolor= blue,
    linkcolor= blue,
    filecolor= blue,      
    urlcolor= blue,
    pdfauthor = {Chuan Chen},
    pdfcreator = {\LaTeX\ and \flqq hyperref\frqq},
}

\usepackage[capitalise]{cleveref}

\graphicspath{ {./Figs/} }

\newcommand{\Fref}[1]{Fig.~\ref{#1}}

\newcommand{\Eqref}[1]{Eq.~\eqref{#1}}
\newcommand{\Eqrefs}[1]{Eqs.~\eqref{#1}}
\newcommand{\Secref}[1]{Sec.~\ref{#1}}

\newcommand{\Appref}[1]{Appendix~\ref{#1}}

\newcommand{\MCS}{\mathrm{MCS}}
\newcommand{\vor}{\mathrm{vor}}
\newcommand{\meV}{\,\text{meV}}

\newcommand{\p}{\mathrm{p}}
\newcommand{\dm}{\mathrm{d}}
\newcommand{\br}{\bm{r}}
\newcommand{\bk}{\bm{k}}
\newcommand{\bq}{\bm{q}}

\newcommand{\bAh}{\bm{A}^h}
\newcommand{\bAs}{\bm{A}^s}
\newcommand{\bAe}{\bm{A}^e}
\newcommand{\bEh}{\bm{E}^h}
\newcommand{\bEs}{\bm{E}^s}
\newcommand{\bEe}{\bm{E}^e}
\newcommand{\barAh}{\bar{A}^h}
\newcommand{\bjh}{\bm{j}^h}

\makeatletter
\newcommand*{\rom}[1]{\expandafter\@slowromancap\romannumeral #1@}
\makeatother

\DeclareMathOperator{\sgn}{sgn}

\begin{document}
\title{Non-Ioffe-Larkin composition rule and spinon-dictated electric transport in doped Mott insulators}

\author{Chuan Chen}
\affiliation{
Lanzhou Center for Theoretical Physics,
Key Laboratory of Quantum Theory and Applications of MoE,
Key Laboratory of Theoretical Physics of Gansu Province,
and School of Physical Science and Technology,
Lanzhou University, Lanzhou, Gansu 730000, China
}
\affiliation{
Institute for Advanced Study, Tsinghua University,
Beijing 100084, China
}

\author{Jia-Xin Zhang}
\email[Correspondence to: ]{zhangjx.phy@gmail.com}
\affiliation{
Institute for Advanced Study, Tsinghua University,
Beijing 100084, China
}

\author{Zhi-Jian Song}
\affiliation{
Institute for Advanced Study, Tsinghua University,
Beijing 100084, China
}

\author{Zheng-Yu Weng}
\affiliation{
Institute for Advanced Study, Tsinghua University,
Beijing 100084, China
}

\begin{abstract}
The electric resistivity is examined in the constrained Hilbert space of a doped Mott insulator, which is dictated by a non-Ioffe-Larkin composition rule due to the underlying mutual
Chern-Simons topological gauge structure. In the low-temperature pseudogap phase, where holons remain condensed while spinons proliferate, the charge transport is governed by a chiral
spinon excitation, comprising a bosonic spin-$1/2$ at the core of a supercurrent vortex. It leads to a vanishing resistivity with the ``confinement'' of the spinons in the 
superconducting phase but a low-$T$ divergence of the resistivity once the spinon confinement is disrupted by external magnetic fields. In the latter, the chiral spinons will generate
a Hall number $n_H =$ doping concentration $ \delta$ and a Nernst effect to signal an underlying long-range entanglement between the charge and spin degrees of freedom.
Their presence is further reflected in thermodynamic quantities such as specific heat and spin susceptibility. Finally, in the high-temperature spin-disordered phase,
it is shown that the holons exhibit a linear-$T$ resistivity by scattering with the spinons acting as free local moments, which generate randomized gauge fluxes as perceived by
the charge degree of freedom.
\end{abstract}

\date{\today}

\maketitle

%
\section{Introduction}
The characteristics of a correlated state of matter, including the nature of its elementary excitations, are often reflected in its transport properties.
In high-$T_c$ cuprates, different phases in their phase diagram exhibit diverse behaviors of electric resistivity $\rho^e$:
(i)~Near half-filling, the system is an antiferromagnetic (AFM) Mott insulator with charge localization, which can be quickly destroyed by doping;
(ii)~At high temperatures, the finite-doped system is in a strange metal (SM) phase with $\rho^e \propto T$ extending beyond the Mott-Ioffe-Regel
limit~\cite{Hussey.Cooper.2009hpf, Phillips-Abbamonte2022, Gurvitch1987, Takagi1992, Campuzano.Kaminski.2003};
(iii)~As the system enters the pseudogap (PG) regime at lower temperatures, $\rho^e$ starts to deviate from the linear-$T$ behavior,
resembling a partial depletion of the low-lying charge carriers' density of states~\cite{Takagi1992, Barisic2013, Timusk1999};
(iv)~The resistivity vanishes at the low-temperature superconducting (SC) transition
but can become insulating when strong external magnetic fields suppress SC condensation~\cite{Ando1995,Boebinger1996,Ando2004},
although some recent works suggest a metal-like finite upturn at
$T\rightarrow 0$~\cite{Taillefer.Proust.2019,Daou2008, Badoux2016, Doiron-Leyraud2017};
(v)~Near a critical doping $\delta^*$, the PG phase terminates
and an SM phase with $\rho^e \propto T$ extends down to much
lower temperatures~\cite{Taillefer.Proust.2019}.
Concurrently, the Fermi liquid (FL) phase with $\rho^e \propto T^2$ emerges and strengthens with increasing 
doping $> \delta^*$~\cite{Hussey.Cooper.2009hpf}.
\begin{figure}
\centering
\includegraphics[width=\linewidth]{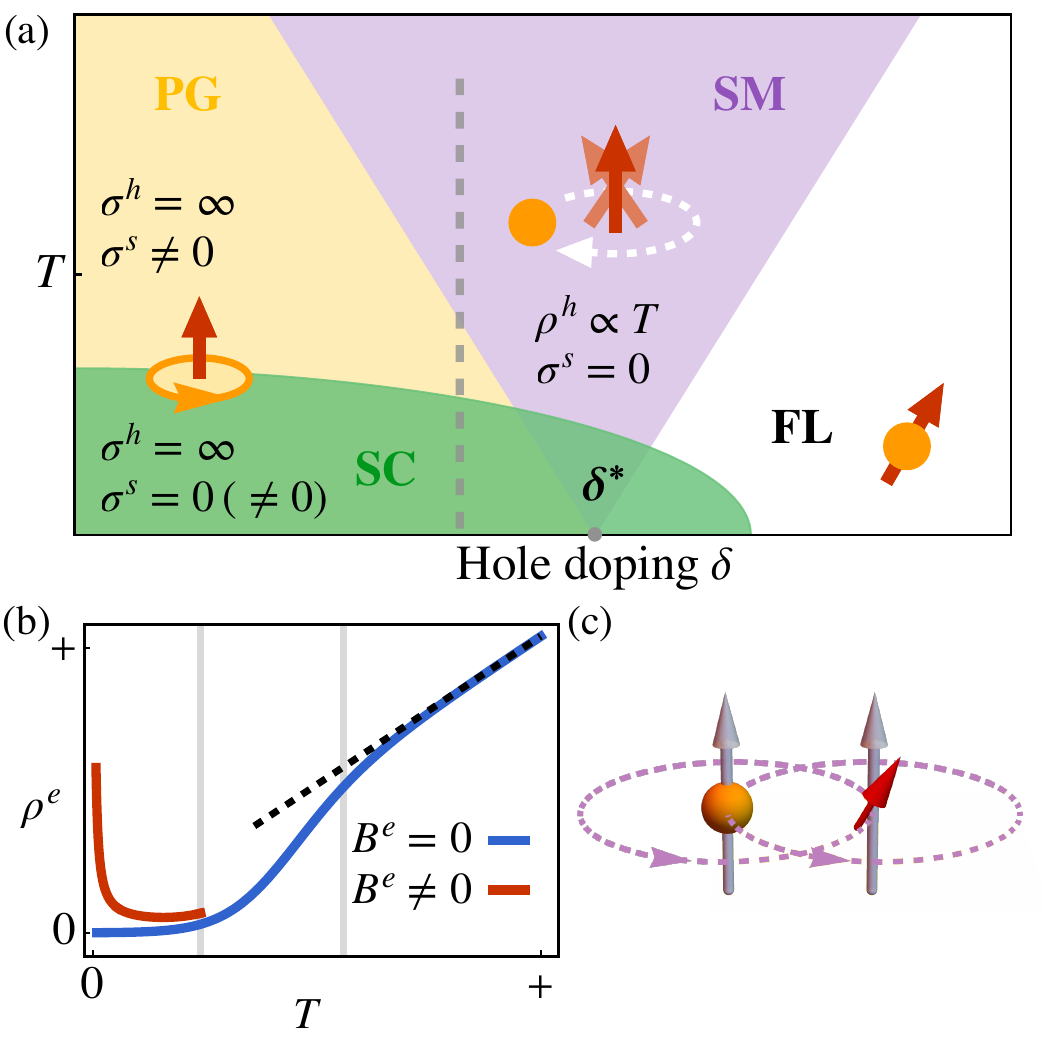}
\caption{(a) Summary of conductance behaviors: $\sigma^s$ for spinons (red and yellow arrows) and $\sigma^h$ for holons (yellow dots),
in the phase diagram of doping and temperature. 
The nonzero $\sigma^s$ in the parenthesis indicates the situation when SC is suppressed by magnetic fields.
(b) Behavior of electrical resistivity 
across temperature regions along the gray dashed line highlighted in (a).
(c) The mutually-seen $\pi$-flux tubes attached to spinons and holons in the phase-string framework.}
\label{fig:pd}
\end{figure}
Such complex phenomena are difficult to fit into the FL theory, where electric resistivity is attributed to dressed electrons/holes.
The challenges stem from the significant influence of strong on-site Coulomb repulsion, which imposes a no-double-occupancy (NDO)
constraint: $ \sum_{\sigma} c_{i,\sigma}^\dagger c_{i,\sigma} \le 1$,
where $c_{i,\sigma}$ is the electron annihilation operator.
Within this low-energy subspace, each Cu-O plane can be effectively described by a single-band $t$-$J$ model~\cite{Lee-Wen2006}.
However, understanding the complex phenomena in cuprates, including their transport properties, from the microscopic $t$-$J$ model 
remains challenging due to the strong correlation effect between the spin and charge degrees of freedom inherited from the NDO constraint.

A promising approach for studying the $t$-$J$ model and handling the NDO constraint is the renowned parton
construction~\cite{Lee-Wen2006}.
In this paper, we discuss the electrical transport behaviors of the $t$-$J$ model
derived from the phase-string theory~\cite{Sheng-Weng1996,Weng-Ting1997},
which incorporates a mutual Chern-Simons (MCS) topological gauge structure that naturally implements the NDO constraint~\cite{Kou-Weng2005}.
The resulting non-Ioffe-Larkin composition rule systematically describes distinct behaviors of electric resistivity 
across different phases, as summarized in \Fref{fig:pd}(a)-(b), aligning with experimental results.
In particular, when holons condense at low temperatures in the regime of $\delta < \delta^*$, the transport will be solely determined by charge-neutral spinons.
These spinon excitations are able to capture the magnetic-field-induced SC-insulator transition, predict a Hall number $n_H \propto \delta$,
and yield a Nernst signal that aligns closely with experimental data~\cite{Badoux2016,Collignon-Taillefer2017,Lizaire2021, Wang-Ong2003, Ong.Wang.2001, Hardy.Wang.2002}.
The presence of these spinon excitations is also evidenced in the thermodynamic observables such as specific heat and magnetic susceptibility.
Finally, the spinons are shown to provide the strongest scattering mechanism for the charge transport in the SM regime.

\section{Ioffe-Larkin composition rule in slave-boson theory}
\begin{figure}
    \centering
    \includegraphics[width = \linewidth]{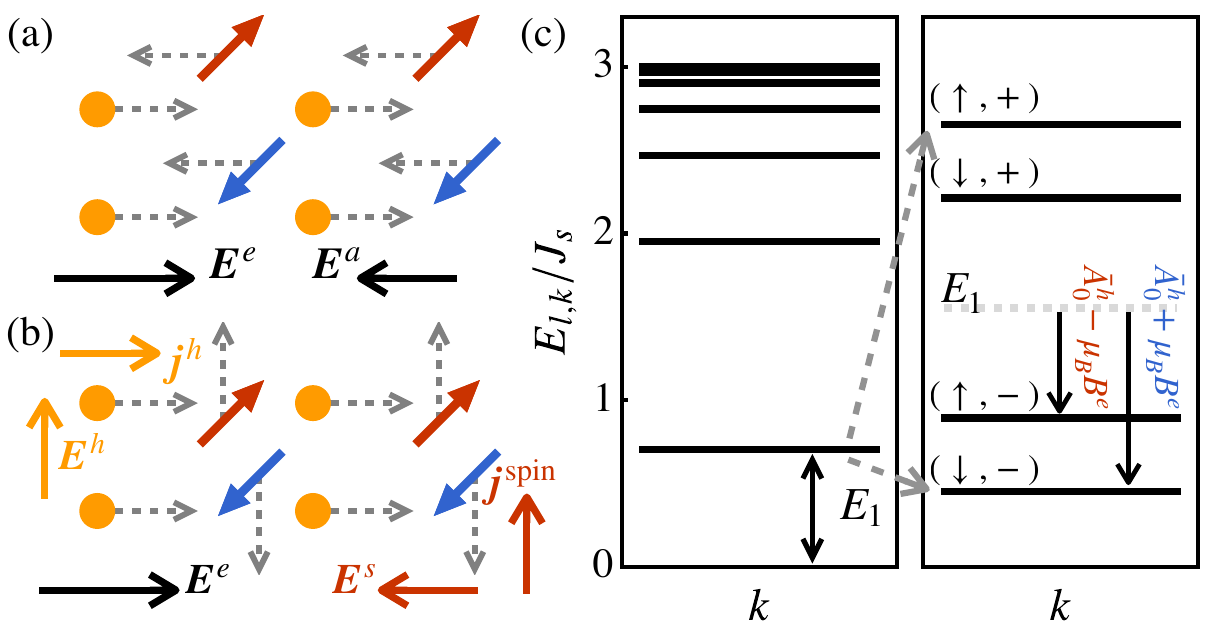}
    \caption{(a) The holon (yellow dots) and backflow spinon (red/blue arrows) currents in a $U(1)$ slave-boson theory.
    (b) In phase-string theory, a holon (spin) current generates a transverse $\bm{E}^h$ ($\bm{E}^s$) field perceived by spinons (holons) due to the flux attachements.
    (c) Mean-field energy levels of $b$-spinons.
    Each degenerate energy level $E_l$ at $B^e = 0$ is split at a finite $B^e$ into $E_l \pm \bar{A}^h_0 \pm \mu_B B^e$.}
    \label{fig:IL}
\end{figure}

We start by briefly reviewing the Ioffe-Larkin composition rule in a conventional parton theory, taking the example of the $U(1)$ slave-boson theory
(a detailed derivation can be found in \Appref{sec:Ioffe-Larkin}).
In this case~\cite{Lee-Wen2006,Baskaran1988,Ioffe-Larkin1989,Lee-Nagaosa1992}, each electron is fractionalized into a (charged) bosonic holon $h_i$ and
a (spinful) fermionic spinon $f_{i,\sigma}$:
$c_{i,\sigma} \leftrightarrow h_i^\dagger f_{i,\sigma}$,
and the NDO constraint is replaced by a holon/spinon single-occupancy condition:
\begin{equation} \label{eq:SO-cons}
    h_i^\dagger h_i + \sum_\sigma f_{i,\sigma}^\dagger f_{i,\sigma} = 1,
\end{equation}
such that spinons and holons cannot occupy the same site.
At low energies, the system can be described by a $U(1)$ gauge theory with both types of matter coupled to emergent
gauge fields $a_\mu$~\cite{Baskaran1988}. Integration over $a_\mu$ results in \Eqref{eq:SO-cons} with a
cancellation of holon and spinon currents:
$\bm{j}^h = - \bm{j}^s $, i.e., the holons' movement is always accompanied by a spinon backflow,
as illustrated in \Fref{fig:IL}(a).
This leads to the so-called Ioffe-Larkin composition rule~\cite{Ioffe-Larkin1989,Ioffe-Kotliar1990, Nagaosa-Lee1990, Nagaosa-Lee1992,Lee-Nagaosa1992}:
\begin{equation} \label{eq:Ioffe-Larkin}
    \rho^e = \rho^h + \rho^s.
\end{equation}
Here $\rho^h$ and $\rho^s$ denote respectively the resistivities of holons and spinons. 

\section{Phase-string theory and the non-Ioffe-Larkin composition rule}
Alternatively, instead of the $U(1)$ gauge fluctuation in the slave-boson scheme, the NDO constraint in the $t$-$J$ model
can also be implemented via a flux-attachment treatment.
In the phase-string theory, each holon ``carries'' a $\pi$-flux tube perceived by spinons,
and \emph{vice versa} (see \Fref{fig:pd}(c) for an illustration)~\cite{Weng-Ting1997,Kou-Weng2005}.
When holons (spinons) condense, the bound flux tube from each spinon (holon) 
induces a charge (spin) vortex, ensuring the condensate is excluded from the vortex core sitting by the opposite
species to maintain ~\Eqref{eq:SO-cons}.
This scenario is captured by a MCS gauge theory with the Lagrangian
$L = L_h + L_b + L_{\MCS}$~\cite{Kou-Weng2005,Qi-Weng2007,Weng2011}, where


\begin{subequations}
\begin{align}
L_h = & \sum_I h_I^{\dagger} (\partial_\tau-i A_{0}^s -i  A_{0}^e + \mu_h ) h_I \nonumber \\
& -t_h \sum_{I,\alpha} ( h_{I+\hat{\alpha}}^{\dagger} h_{I} e^{i A_{\alpha}^s(I)
+ i A_{\alpha}^e(I)} + h.c. ), \label{eq:L_h} \\
L_b = & \sum_{i,\sigma} b_{i,\sigma}^{\dagger} ( \partial_\tau-i \sigma A_0^h + \lambda_b
+ \sigma \mu_B B^e ) b_{i,\sigma} \nonumber \\
& -\frac{J_{\text {eff }}}{2} \Delta_s \sum_{i, \alpha, \sigma} ( b_{i+\hat{\alpha}, \sigma}^{\dagger} b_{i, -\sigma}^{\dagger} 
e^{i \sigma A_{\alpha}^h (i)}+ h.c. ), \label{eq:L_b} \\
L_{\MCS} = & \frac{i}{\pi} \sum_i \varepsilon_{\mu \nu \lambda} A_\mu^s(I) 
\partial_\nu A_\lambda^h (i) \label{eq:L_CS}.
\end{align}
\end{subequations}
Here indices $i$ and $I$ represent the two-dimensional square lattice site and its dual lattice site, respectively.
Microscopically, the MCS topological gauge structure originates from a nontrivial sign structure encoded in the $t$-$J$
model, as thoroughly discussed in Refs.~\cite{Sheng-Weng1996,Weng-Ting1997,Wu-Zaanen2008,Lu-Weng2023,PhysRevLett.133.126501, ZhangZhang-Weng2024}.
Unlike the conventional $U(1)$ slave-boson theory, a key feature of $L$ above is that each holon $h$ (spinon $b$) 
is attached to a $\pi$ ($\pm \pi$ depending on spin-$\sigma$) flux tube of $A^h$ ($A^s$), which is coupled to spinons (holons), as indicated by
the equation of motions for $A_0^s$ and $A_0^h$:
\begin{equation}\label{eq:con-MCS}
    \pi n_I^h = \nabla \times \bAh, \quad \pi \sum_\sigma \sigma n_{i \sigma}^b = \nabla \times \bAs.
\end{equation}
Note here both holons and spinons are \emph{bosons}, with the restoration of fermionic statistics in the composite particles.


Similar consideration for $\bAs$ ($\bAh$) implies that the holon (spin) current
$j^{h/\mathrm{spin}}_\alpha = -\partial L_{h/b}/\partial A^{s/h}_\alpha$
is associated with an ``electric'' field $E_\alpha^{h / s}=i ( \partial_\alpha A_0^{h / s}-\partial_0 A_\alpha^{h / s} )$:
\begin{equation} \label{eq:j-E}
j^{h/\mathrm{spin}}_\alpha = \frac{1}{\pi} \varepsilon_{\alpha \beta} E^{h/s}_\beta,
\end{equation}
consistent with the fact that the movement of ``magnetic'' fluxes will generate an ``electric'' field.
Combining with 
\begin{equation}\label{eq:js}
    \bjh = \sigma^h (\bEs + \bEe), \;\;\;\; \boldsymbol{j}^\mathrm{spin} = \sigma^s \bEh
\end{equation}
one obtains $j^\mathrm{spin}_\alpha = -\pi \sigma^s \varepsilon_{\alpha \beta} j^h_\beta$ where $\varepsilon_{xy} = 1 = -\varepsilon_{yx}$
is the anti-symmetric tensor~\cite{Ye-Weng2011,Ye-Weng2012}.
For diagonal $\sigma^{s}$ and $\sigma^h$, the holon and spin currents are perpendicular to each other, as illustrated in \Fref{fig:IL}(b).
This contrasts with the back-flow picture in $U(1)$ slave-boson theories (see \Fref{fig:IL}(a)).
Combining \Eqref{eq:j-E} and \Eqref{eq:js}, one obtains:
\begin{equation}\label{eq:j^h}
    \bm{j}^h = - \pi^2 \sigma^h \sigma^s \bm{j}^h + \sigma^h \bm{E}^e,
\end{equation}
thus the resistivity reads~\footnote{Here we have used the fact that $\sigma^s_{xx} = \sigma^s_{yy}$ and 
$\sigma^s_{xy} = -\sigma^s_{yx}$ due to the 4-fold rotation symmetry of a square lattice.}:
\begin{equation} \label{eq:rho-PS}
\rho^e = \rho^h+\pi^2 \sigma^s.
\end{equation}
Here we have set $\hbar = 1 = e$~\footnote{Note that $\rho^h$ and $\sigma^s$ are defined by setting
their gauge charge to be $1$, so $[ \rho^h ] = \hbar$, $[ \sigma^s ] = \hbar^{-1}$. As $[ \rho^e ] = \hbar/e^2$
in $2D$, after putting back $\hbar$ and $e$, the full formula of $\rho^e$ reads:
$\rho^e = \hbar/e^2 ( \rho^h/\hbar + \pi^2 \hbar \sigma^s)$.
}.
The contribution of $\sigma^s$ to $\rho^e$ arises from the fact that the $\bm{j}^\mathrm{spin}$-induced $\bm{E}^s$
acts to ``screen'' external $\bm{E}^e$
(note their opposite direction in \Fref{fig:IL}(b)).



Interestingly, when combined with mean-field parameters in $L$ and the corresponding phase diagram~\cite{Ma-Weng2014,Weng2011},
\Eqref{eq:rho-PS} offers a self-consistent view of the transport properties of doped cuprates:
(i) in the AFM phase, $b$-spinons are condensed ($\sigma^s = \infty$) whereas holons are
localized ($\rho^h = \infty$), so the system is insulating;
(ii) in the SC phase, $b$-spinons are gapped ($\sigma^s = 0$) whereas holons are condensed ($\rho^h = 0$), so $\rho^e = 0$;
(iii) above PG in the SM phase, $b$-spinons lose their pairing ($\Delta_s = 0$ in \Eqref{eq:L_b})
and behave as free local moments ($\sigma^s = 0$) inducing randomized
$B^s = \nabla \times \bm{A}^s$ fluxes.
This leads to $\rho^e = \rho^h \propto T$~\cite{Gu-Weng2007};
(iv) in FL, holons and spinons are recombined to form electronic quasiparticles which
produces $\rho^e \propto T^2$~\cite{Zhang.Weng_2023}.
Note that although \Eqref{eq:rho-PS} was introduced previously in Refs.~\cite{Ye-Weng2011,Ye-Weng2012,Ma-Weng2014},
and calculations of $\rho^e$ across various dopings and temperatures were conducted in Ref.~\cite{Ma-Weng2014},
a detailed analysis of the effect of external magnetic fields has been missing.
Therefore in this paper, we focus primarily on the low-temperature PG phase
where holons have a finite condensation amplitude ($\sigma^h \rightarrow \infty$), and
investigate the impact of magnetic fields.

%
\section{Chiral spinons in PG} \label{sec:chiral-spinons}
In PG, $h_r \approx \sqrt{\delta} e^{i\theta^h_r}$, where $\delta = a^2 \rho_0$ is the 
holon number per site and $a$ is the lattice constant.
At low energies, only the phase fluctuation is important. 
The holon part of the Lagrangian (under a continuous approximation) reads:
\begin{align} \label{eq:L_h-PG}
L_h = & \int d^2 r \ i \rho_0 ( \partial_0 \theta^h - A^s_0 - A^e_0 +\lambda_h ) \nonumber \\
    & +\frac{\rho_0}{2 m_h} ( \nabla \theta^h - \bAs- \bAe )^2.
\end{align}
To consider the holons' phase vortices, one can replace $\partial_\mu \theta \rightarrow a_{\mu}$
with $\varepsilon_{\mu \nu \lambda} \partial_\nu a_{\lambda} = 2\pi j^\mathrm{vor}_\mu$, where $j^\mathrm{vor}_\mu$
is the vortex current. Such a constraint can be implemented by introducing an auxiliary field
$\tilde{A}_\mu$, and $\mathcal{L}_h$ reads:
\begin{align}
\mathcal{L}_h = & i \rho_0 ( a^\vor_0 - A^s_0 - A^e_0 +\lambda_h ) 
+\frac{\rho_0}{2 m_h} ( \bm{a}^\vor - \bAs- \bAe )^2 \nonumber \\
& + \frac{i}{\pi} \varepsilon_{\mu \nu \lambda} \tilde{A}_\mu \partial_\nu a^\vor_\lambda
-i 2 \tilde{A}_\mu j^\vor_\mu.
\end{align}
The kinetic energy can be decoupled through a Hubbard-Stratonovich transformation:
\begin{align}
\sum_{\mu = 1,2} \frac{1}{2}\frac{m_h}{\rho_0}J_\mu^2 + i J_\mu (a^\vor_\mu - A^s_\mu - A^e_\mu).    
\end{align}
The integration over $A^s$ enforces: $J_\mu = \frac{1}{\pi} \varepsilon_{\mu \nu \lambda} \partial_\nu A^h_\lambda$,
here we have set $J_0 = \rho_0$. Furthter integrating out $a^\mathrm{vor}$ gives rise to:
$\tilde{A}_{\mu} = -A^h_{\mu} - \partial_\mu \Lambda$.
Finally, $\mathcal{L}_h$ can be recast into:
\begin{align} \label{eq:L_h-dual}
\mathcal{L}_h = & \frac{1}{2 \pi^2}\frac{m_h}{\rho_0} [ (\partial_2 A^h_0 - \partial_0 A^h_2)^2 +
(\partial_0 A^h_1 - \partial_1 A^h_0)^2 ] \nonumber \\
& -\frac{i}{\pi} \varepsilon_{\mu \nu \lambda} A^h_\mu \partial_\nu A^e_\lambda
+ i 2 A^h_\mu j^\vor_\mu.
\end{align}
Therefore, $A^h$ couples to three types of ``matters'': holon vortices (with each $2\pi$-vortex
carrying gauge charge $-2$), $b_\sigma$-spinons (with each carrying charge $\sigma$), 
and electromagnetic fluxes (with every $\pi \equiv h/(2e)$ flux carrying charge $1$).
Since $A^h_0$ induces a logarithmic interaction between these charged ``particles'',
it is legitimate to consider only those with smallest gauge charges at low energies.
For the spinons, besides the bare $b_{\uparrow/\downarrow}$-spinon with charge $\pm 1$, the ``fusion''
of a $\pm 2\pi$ holon vortex and a $b_{\uparrow/\downarrow}$-spinon 
has also gauge charge $\mp 1$~\cite{Weng-Qi2006}.
We therefore include 4 types of ``elementary'' particles, to simplify the notation,
they are denoted as $b_{\sigma,v}$ ($\sigma, v = \pm1$)
with $v$ standing for its $A^h$ gauge charge (referred to as \emph{vorticity} henceforth).
The spinons' Lagrangian thus reads:
%
%
\begin{align} \label{eq:new-L_b}
L_b = & \sum_{i,\sigma,v} b_{i,\sigma,v}^{\dagger} ( \partial_\tau-i v A_0^h + \lambda_b + \sigma \mu_B B^e ) b_{i,\sigma,v} \nonumber \\
& -J_s \sum_{i, \alpha, \sigma,v}  b_{i+\hat{\alpha}, \sigma,v}^{\dagger} b_{i, -\sigma,-v}^{\dagger} e^{i v A_{\alpha}^h (i)}+ h.c.
\end{align}
The $\bm{j}^\mathrm{spin}$ in \Eqref{eq:js} should now be replaced 
by the vorticity current $\bm{j}^{\mathrm{v}}$.
Moreover, the $A^h$ charge neutral condition implies:
\begin{equation} \label{eq:charge-neutral}
\sum_{\sigma,v} v b_{i,\sigma,v}^\dagger b_{i,\sigma,v} + a^2 B^e/\pi = 0.
\end{equation}
At mean-field level, from \Eqref{eq:con-MCS}, condensed holons produce a finite $B^h = \pi \delta$ perceived by
$b$-spinons, so the spectra of $b$-spinons' Bogoliubov quasiparticles (Bogolons) are flat Landau levels (LLs)
with non-zero Chern numbers.
When $B^e \neq 0$, \Eqref{eq:charge-neutral} entails a non-zero mean-field value of $A^h_0$,
resulting in a separation of states with opposite $v$ which are degenerate when
$B^e = 0$, as illustrated in \Fref{fig:IL}(c).

%
%
\begin{figure}
\centering
\includegraphics[width= \linewidth]{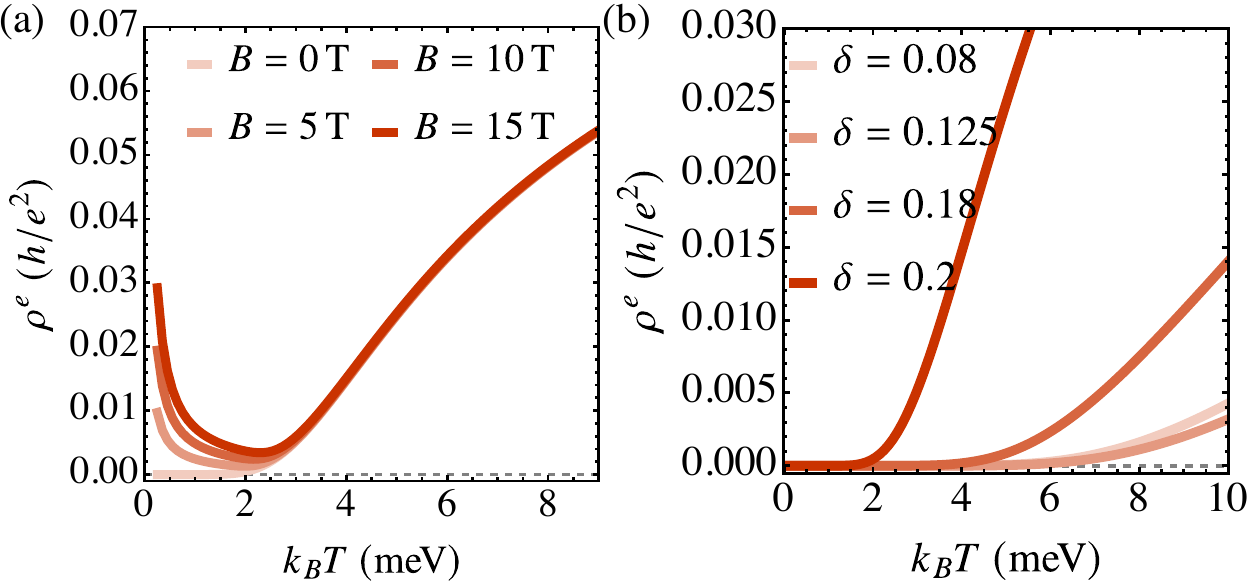}
\caption{
(a) $\rho^e_{xx}$ at $\delta = 0.2$ with different magnetic fields.
At $B = 0\,\mathrm{T}$, the system enters SC at low temperatures ($\rho^e_{xx} \rightarrow 0$);
when SC is suppressed by a finite $B$ field, $\rho^e_{xx}$ shows an insulating behavior with
$\rho^e_{xx} \propto 1/T$.
(b) $\rho^e_{xx}$ at different dopings with $B = 0\,\mathrm{T}$.
}\label{fig:rho_xx}
\end{figure}

\begin{figure}
\centering
\includegraphics[width= \linewidth]{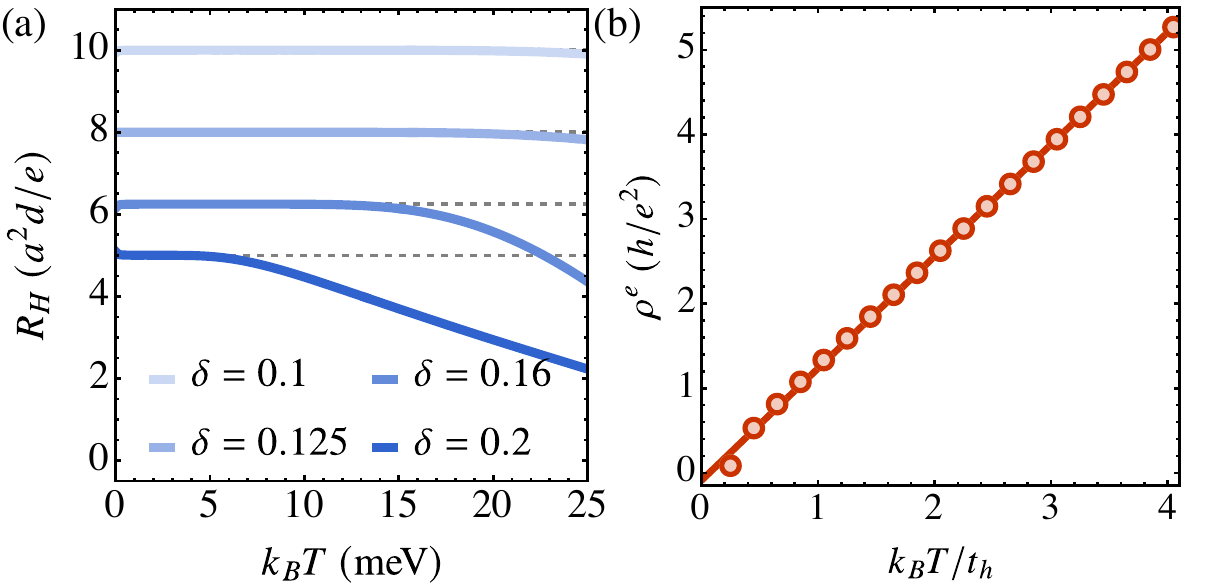}
\caption{
(a) Hall coefficient $R_H$ at different dopings.
At low temperatures, $R_H e/(d a^2)$ saturates to $1/\delta$ (indicated by the
dashed lines), i.e., $n_H = \delta$.
(b) Resistivity from holons scattering with the random flux tubes associated with the spinons as free local moments at high temperatures.
}\label{fig:Hall}
\end{figure}

\subsection{Electrical resistivity in PG}
Now we explicitly consider the low-temperature PG phase where the holons condense with
the effect of the vortex-like phase fluctuation outlined above. 
Here the condensation of holons results in $\rho^h = \frac{\omega}{i}m_h/\rho_0 \rightarrow 0$ in the DC limit.
Based on \Eqref{eq:rho-PS}, the DC electric resistivity in PG is solely determined by $b$-spinons'
conductivity:
\begin{equation} \label{eq:rho^e_LPP}
   \rho^e = \pi^2 \sigma^s,
\end{equation}
with $\sigma^s$ being interpreted as the conductivity of the newly defined 4-component $b$-spinon-vortices.
A detailed derivation of above equation can be found in \Appref{sec:PS-conductivity}.

\begin{figure}
\centering
\includegraphics[width = 0.9\linewidth]{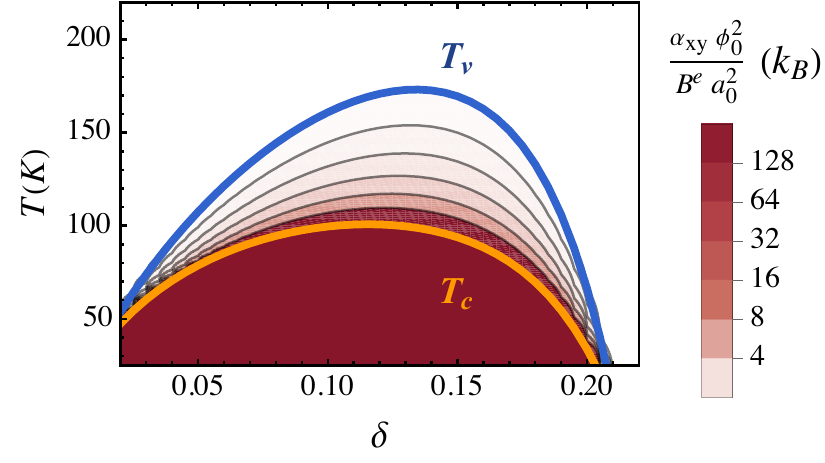}
\caption{
The magnitudes of the Nernst effect as a function of temperature $T$ and doping density $\delta$.\
The orange line indicates the SC transition temperature, $k_B T_c \approx E_g/6.4$, while the blue line represents the temperature $T_v$, at which the Nernst effect vanishes.
}\label{fig:Nernst}
\end{figure}

In the absence of a background magnetic field,
the gapped $b$-spinons (the lowest LL has an energy $E_1 = E_g/2$, with $E_g$ being the
excitation energy of the spin resonate mode~\cite{Mei-Weng2010,Chen-Weng2005})
results in $\sigma^s_{xx} \rightarrow 0$ at low temperatures. The superconducting critical temperature $T_c \propto E_g$,
consistent with experimental findings~\cite{Mei-Weng2010}, is shown by the orange line in \Fref{fig:Nernst},
where the magnitude obtained from our mean-field framework exhibits a dome-shaped evolution as a function of doping. 
\Fref{fig:rho_xx}(b) shows $\rho^e_{xx}$ at various dopings without $B^e$, where
$T_c$ depends on $\delta$.
In calculating $\rho^e_{xx}$, we have introduced a broadening factor $\Gamma = 0.02 J \ll E_g$
(with $J = 120 \meV$) into $b$-Bogolons' spectral function to account for fluctuations
beyond the mean-field theory (and extrinsic contributions like disorder).
Different choices of $\Gamma$ do not alter the qualitative behavior of the results~\cite{Ma-Weng2014}.

On the other hand, SC can be killed by magnetic field, and the system becomes an insulator with
$\rho^e_{xx} \propto 1/T$ at low temperatures.
This is because \Eqref{eq:charge-neutral} enforces excitement of $b$-spinon Bogolons
with an amount proportional to $B^e$.
Such a magnetic field induced SC-insulator transition agrees qualitatively with experimental
findings, although $\rho^e \propto \ln(1/T)$ was observed~\cite{Ando1995,Boebinger1996,Ando2004}.

\subsection{Hall coefficient}
In the presence of $B^e$, $b$-spinons also has a finite $\sigma^s_{xy}$ due to their non-trivial
band topology and net vorticity. It can be shown that (see \Appref{sec:b-spinon} for more details):
\begin{align}
\sigma^s_{xy} =& \sum_{l,\sigma,v} \int \frac{d^2k}{(2\pi)^2} \, v \,
\mathcal{F}^l_{xy,k} n_B(E_{l,k,\sigma,v}) \nonumber \\
\approx & \frac{1}{2\pi} \sum_{l,\sigma,v}  \mathcal{C}_l \, v \,
n_B(E_{l,\sigma,v}).
\end{align}
Here $n_B$ is the Bose-Einstein distribution function, $\mathcal{F}^l_{xy,k}$ is the Berry
curvature of the $l$-th LL with Chern number $\mathcal{C}_l$.
In the second line above, we have used the fact that $b$-Bogolons' LLs have negligible $k$-dependence. Plots of the temperature dependence of the Hall coefficient, \( R_H \equiv \rho^e_{yx} d / B^e \) (with \( d \) denoting the distance between adjacent Cu-O layers), at different doping densities are shown in \Fref{fig:Hall} (a). These plots exhibit a plateau at low temperatures and suppressed signals as the temperature increases, in agreement with experimental observations~\cite{Badoux2016}.

Moreover, at low temperatures, only the two lowest LLs ($E_{1,\sigma,v} = E_{N_L,\sigma,v}$,
$N_L \approx 2/\delta$ is the number of LLs) with $v = -1$ have significant occupation,
\Eqref{eq:charge-neutral} implies 
$\sum_\sigma n_B(E_{1,\sigma,-1}) \approx \frac{B^e a^2}{\delta \pi}$.
Since $\mathcal{C}_1 = 1 = \mathcal{C}_{N_L}$, a direct implication is that the Hall coefficient $R_H \equiv \rho^e_{yx} d/B^e = \frac{a^2 d}{e \delta}$,
where $d$ is the distance between adjacent Cu-O layers. 
Therefore the Hall number $n_H = \delta$ within PG, consistent with experimental
findings~\cite{Badoux2016,Collignon-Taillefer2017,Lizaire2021}.





\subsection{Nernst effect}\label{Nernst}
The implications of $b$-spinons for the Nernst effect can be analyzed in a similar manner to Ref.~\cite{Weng-Qi2006}.
A temperature gradient along the $x$-direction $\partial_x T$ can generate a drift motion of $b$-spinons with velocity $v_x^b$ satisfying:
$s_\phi \partial_x T=-\eta_s v^b_x$,
here $s_\phi$ denotes the transport entropy of each $b$-spinon and $\eta_s$ is its viscosity, which depends on the broadening (lifetime) of the spinon spectral function.
As discussed before, a finite $B^e$ will polarize $b$-spinons' vorticity and induce a vortex current:
$\bm{j}^\mathrm{v} = ( n^b_{v=1} - n^b_{v=-1} ) \bm{v}^b$. 
Replacing $\bm{j}^{\mathrm{spin}} \rightarrow \bm{j}^{\mathrm{v}}$ in \Eqref{eq:j-E} and using $\bm{E}^s \rightarrow -\bm{E}^e$,
it can be seen that $j^\mathrm{v}_x$ ``induces'' a perpendicular electric field $E^e_y = -\pi j^\mathrm{v}_x$.
The Nernst signal thus reads~\cite{Weng-Qi2006}: $e_N=E^e_y/|\partial_x T| = B^e s_\phi/\eta_s$.
The viscosity $\eta_s$ is actually also related to spinons' vorticity conductivity $\sigma^s$.
Note $\bm{E}^{h}$ prompts spinons of opposite vorticity to drift in opposite directions:
$\pm \bm{v}^b$ for $v = \pm1$ with $\bm{E}^h = \eta_s \bm{v}^b$, and $\bm{j}^\text{v} = n^b_{\nu=1} \bm{v}^b - n^b_{\nu=-1} (-\bm{v}^b) = n^b \boldsymbol{v}^b$.
Using $\rho^e = \pi^2 \sigma^s$ in the PG, one can define a coefficient $\alpha_{xy}$ independent of $\eta_s$~\cite{Weng-Qi2006,Hardy.Wang.2002}:
\begin{equation} \label{eq:alpha_xy}
   \alpha_{xy} \equiv \frac{e_N}{\rho^e}=\frac{B^e s_\phi}{\Phi_0^2 n^b}.
\end{equation}
Unlike conventional BCS superconductors, the vortex core here
captures a free spin-$1/2$ magnetic moment ($b$-spinon),
thereby contributing to a transport entropy
$s_\phi=k_B\{\ln[ 2 \cosh(\beta \mu_B B^e) ] - \beta  \mu_B B^e
\tanh(\beta \mu_B B^e) \}$~\cite{Hardy.Wang.2002,Weng-Qi2006}.
\Fref{fig:Nernst} shows the temperature and doping density dependence of \( \alpha_{xy}/B^e \), where the temperature \( T_v \), at which the Nernst effect vanishes, is significantly higher than the superconducting critical temperature \( T_c \). This feature aligns quantitatively well with experimental data~\cite{Wang-Ong2003, Ong.Wang.2001, Hardy.Wang.2002}, which cannot be easily explained by conventional type-II superconductivity theory. In conventional theories, the Nernst effect signal, driven by vortices, is only present within the superconducting phase, as vortices are not well-defined for \( T > T_c \), leading to a rapid suppression of the Nernst signal. Therefore, the observed Nernst signal further validates the presence of spinon vortices that carry transport entropy and persist beyond the superconducting order.

\begin{figure}
\centering
\includegraphics[width = \linewidth]{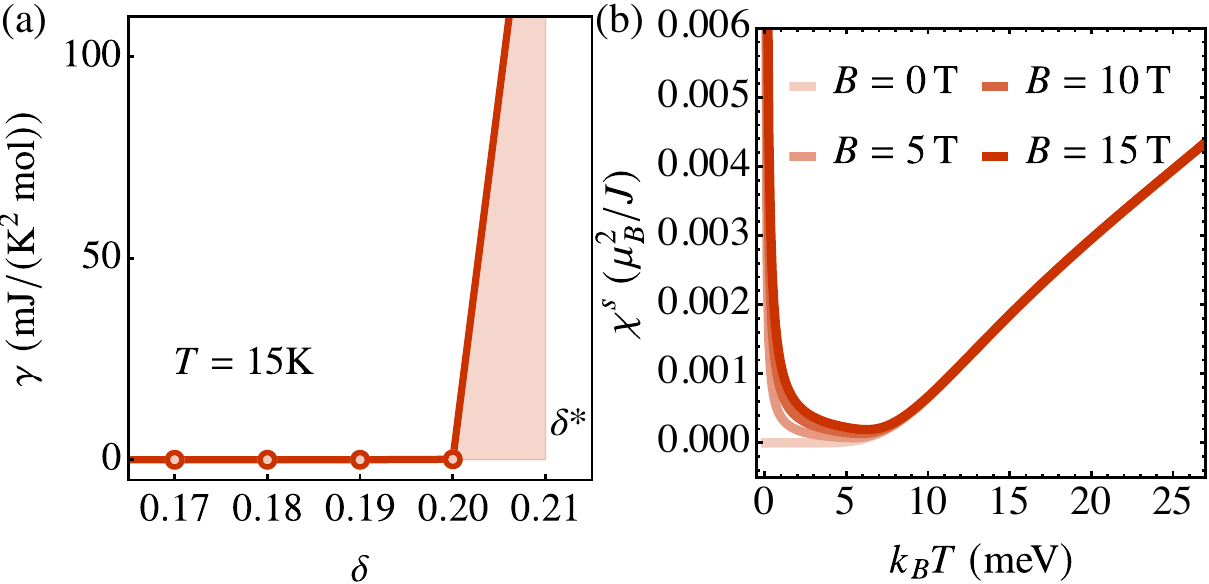}
\caption{
(a) Specific heat contributed by spinons at different dopings.
(b) Uniform spin susceptibility with different magnetic fields at $\delta = 0.2$.
}\label{fig:cV}
\end{figure}

\subsection{Other thermodynamic signatures}
The presence of $b$-spinons is also  reflected in various thermodynamic measurements. 
\Fref{fig:cV}(b) displays the uniform spin susceptibility $\chi^s$ at $\delta = 0.2$, which closely aligns with the electric resistivity shown in \Fref{fig:rho_xx}(a).
When $B^e = 0$,  $\chi^s(T \to 0) = 0$ as $b$-Bogolons are gapped, consistent with standard observations in SC states.
In contrast, when magnetic fields suppress SC coherence, the $B^e$-induced $b$-spinons (from \Eqref{eq:charge-neutral}) acts as free magnetic moments,
resulting in Curie-Weiss behavior $\chi^s \propto 1/T$ at low temperatures. 

Furthermore, the doping dependence of specific heat coefficient $\gamma \equiv C_{V}^{b} / T$ from $b$-spinons,
illustrated in \Fref{fig:cV}(a), exhibits a marked enhancement as doping $\delta \rightarrow \delta^*$ at low temperatures.
This aligns with experimental observations~\cite{Taillefer.Proust.2019, Klein.Girod.2020} and suggests an instability
of the $b$-spinon RVB order at $\delta \approx \delta^*$, marking the breakdown of the PG phase.
The expressions for these quantities are provided in \Appref{sec:heat}.

%
\section{Discussion}
We have explored charge transport in a low-$T$ PG phase within the phase-string description of
the $t$-$J$ model.
This phase is characterized by holon condensation but lacks SC phase coherence due to the strong 
phase fluctuations induced by excited spinons under an MCS gauge structure.
Notably, applying an external magnetic field at low temperatures may stabilize this phase, 
leading to an SC-insulator transition driven by the proliferation of deconfined spinons.

In the MCS theory, the mean-field spinon spectrum consists of flat Landau levels (LLs) with infinite effective mass,
as they perceive a uniform background $A^h$-flux generated by condensed holons.
Consequently, their conductivity $\sigma^s$ (thus $\rho^e$ from \Eqref{eq:rho^e_LPP}) can be vanishingly small even
at finite temperatures, in contrast to the behavior of conventional slave-boson theory.
However, fluctuations in the holon density will lead to variations in the background $A^h$-flux,
inducing tunneling motion of spinons between neighboring cyclotron orbitals.
To capture this fluctuation-induced effect, we have introduced a broadening factor $\Gamma$ in the spinon spectral function,
assuming it to be $T$-independent as a leading-order approximation. 
While $\Gamma$ should generally have a temperature dependence, the qualitative behavior of $\sigma^s$ (and $\rho^e$)
could remain valid even if a $T$-dependent $\Gamma$ is considered. To be more specific, in the absence of an external magnetic field, there is a finite excitation gap $E_1$ for spinons' quasiparticles
(as shown in \Fref{fig:IL}(c)), therefore the Bose-Einstein occupation factor $n_B(E_{1,\downarrow,-1}) \approx e^{-E_{1,\downarrow,-1}/(k_B T)}$
dominates the temperature dependence of $\sigma^s$ (see its expression in \Eqref{eq:sigma^s_xx}) as $T \rightarrow 0$,
leading to $\sigma^s(T \rightarrow 0) \rightarrow 0$. This remains valid even if $\Gamma$ has a power-law dependence on $T$. Also, under an applied magnetic field, the constraint between magnetic flux and spinons' vorticity (see \Eqref{eq:charge-neutral}) requires a finite occupation of spinon quasiparticles, driving $E_{1,\downarrow,-1} \rightarrow 0$ as $T \rightarrow 0$. Since $n_B(E_{1,\downarrow,-1})$ is nearly a constant at low temperatures,  $\Gamma(T)$ becomes crucial in determining the low-$T$ behavior of $\rho^e$. 
If $\Gamma(T)$ remains finite as $T\rightarrow 0$, partial condensation of the $b$-spinons always leads to a divergent $\rho^e(T \rightarrow 0)$,
rendering the system insulating.
However, if $\Gamma(T) \propto T$, $\rho^e$ saturates to a constant, indicating a magnetic-field-induced SC-to-metal transition,
consistent with recent experimental observations~\cite{Daou2008,Doiron-Leyraud2017,Taillefer.Proust.2019}.



Moreover, there are two direct ways to compare with experiments without assuming that $\Gamma$ is temperature-independent.
First, note that the broadening factor in the spinon spectral function is directly related to the width of the resonance peak observed in
neutron scattering experiments~\cite{Dai1999}, a connection established in previous studies~\cite{Zhang.Weng_2024}.
Therefore, the specific temperature dependence of $\Gamma$ can be treated as a phenomenological parameter obtained by fitting the spin spectrum.
Secondly, both Nernst signals $e_N$ and longitudinal resistivity $\rho^e$ depend on this phenomenological parameter (appearing in  $\eta_s$ in \Secref{Nernst}).
However, their ratio -- given by \Eqref{eq:alpha_xy} -- is independent of this parameter and can be directly compared with experimental observables.


The Hall coefficient contributed by the spinon-vortices is $n_H = \delta$.
Free spinons also influence thermodynamic quantities such as specific heat and spin susceptibility.
Their thermal transport properties like the Nernst and thermal Hall
effects above $T_c$ have been previously studied elsewhere~\cite{Song-Weng2023,han2025}.  

Finally, when the short-range RVB pairing of spinons is destroyed by temperature or doping, 
holons will experience an even \emph{stronger} phase fluctuations from the local moments of disordered spins,
behaving like random flux tubes. This leads to strange-metal behavior with resistivity $\rho^e \propto T$
at high-temperatures (see \Fref{fig:Hall}(b))~\cite{Gu-Weng2007}.
On the other hand, a FL phase can emerge at low temperatures when the doping density exceeds $\delta^*$
and RVB pairing vanishes.
Here only the Landau quasiparticles as gauge-neutral ``composite fermions'' formed
by the fusion of fractionalized particles survive the strongest frustration from the random fluxes
of local moments, resulting in $\rho^e \propto T^2$~\cite{Zhang.Weng_2023}.
It is worth noting that some experimental studies on the quasiparticle lifetime $\tau(\omega,T)$, 
based on magnetoresistance~\cite{Proust2016-yd} and optical conductivity~\cite{Mirzaei2013-pf},
have observed FL-like behavior even in the PG phase.

\acknowledgments
We acknowledge stimulating discussions with Bo Li, Jisi Xu and Zhi-Yuan Yao.
Financial support by MOST of China (Grant No.~2021YFA1402101) and NSF of China (Grant No.~12347107,
No.~12404175 and No.~12247101) is acknowledged.

\appendix

\section{Derivation of Ioffe-Larkin rule for $U(1)$ slave-boson theory} \label{sec:Ioffe-Larkin}

The well-known Lagrangian for $U(1)$ slave boson theory is given by $L^{U(1)}=L_h^{U(1)}+L_f^{U(1)}$, of which $h$ and $f$ denotes the bosonic holon and fermionic spinon, respectively. Their explicit expressions are:

\begin{subequations}
\begin{align}
L_h^{U(1)} = & \sum_i h_i^{\dagger} (\partial_\tau-i a_0 -i  A_{0}^e + \mu_h ) h_i \nonumber \\
& -t_h \sum_{i,\alpha} ( h_{i+\hat{\alpha}}^{\dagger} h_{i} e^{i \boldsymbol{a}_{\alpha}(i)
+ i A_{\alpha}^e(i)} + h.c. ), \label{eq:L_hU1} \\
L_f^{U(1)} = & \sum_{i,\sigma} f_{i,\sigma}^{\dagger} ( \partial_\tau-i a_0 + \lambda_f
+ \sigma \mu_B B^e ) f_{i,\sigma} \nonumber \\
& -\frac{J_\mathrm{eff}}{2} \Delta_s\sum_{i, \alpha, \sigma} ( f_{i+\hat{\alpha}, \sigma}^{\dagger} f_{i, -\sigma}^{\dagger} 
e^{i \boldsymbol{a}_{\alpha} (i)}+ h.c. ) \label{eq:L_bU1},
\end{align}
\end{subequations}
where $t_h$ and $\mu_h$ is the hopping integral and chemical potential of holons, while $\Delta_s$ and $\lambda_f$ are the pairing term and chemical potential of spinons.
As shown in \Eqref{eq:L_h} and \Eqref{eq:L_b} of the main text, besides the coupling to external electromagnetic
 $A_{\mu}^e$ (with $\mu=\{ \tau,x,y \}$ containing all the time-space components),
 the basic interplay between holons and spinons are the emergent internal $U(1)$ gauge field $a_{\mu}$, which arise to implement the NDO. The spatial components of $a_{\mu}$ give the constraint between holon current $\boldsymbol{j}^h$ and spinon current $\boldsymbol{j}^f$:
 \begin{equation}\label{back1}
 \frac{\partial L^{U(1)}}{\partial \boldsymbol{a}(i)}=0 \Longrightarrow \boldsymbol{j}^h(i)+\boldsymbol{j}^f(i) =0,
 \end{equation}
 which corresponds to the backflow effects, as shown in Fig.~2(a), indicating that holons moving forward will always push spinons backward. Such induced spinon current will further generate an internal ``electric field'' 
 \begin{equation}\label{back2}
    \boldsymbol{E}^a (i)= \boldsymbol{j}^f (i)/\sigma^f,
 \end{equation}
 where $\sigma^f$ denotes the spinon conductance. 
 In the presence of an external electric field $\boldsymbol{E}^e$, the total field perceived by holon is $\boldsymbol{E}^a + \boldsymbol{E}^e$, leading to the relation between electric (holon) current $ \boldsymbol{j}^e$ ($\boldsymbol{j}^h$) and holon conductance $\sigma^h$:
  \begin{equation}\label{back3}
    \boldsymbol{j}^e (i)=\boldsymbol{j}^h (i)=\sigma^h (\boldsymbol{E}^a + \boldsymbol{E}^e).
 \end{equation}
 Combining with \Eqrefs{back1}-(\ref{back3}), one can obtain the generic series relation for the resistivity $\rho^e$, i.e., Loffe-Larkin rule, as follows:
 \begin{equation}\label{ILrule}
     \rho^e=\rho^h+\rho^f,
 \end{equation}
 where $\rho^h$ and $\rho^f$ denote that resistivity contributed from holons and spinons.
 
The $\sigma^{h/f}$ can be obtained by evaluating partons' current-current correlation functions based on the mean-field Hamiltonian.
For further details and discussions, we refer readers to section IX.~A of Ref.~\cite{Lee-Wen2006}, particularly around Eq.~(63).

\section{Mean-field phase diagram of phase-string theory} \label{sec:mean-field}
In this section, we provide additional details about the mean-field parameters used in the main text. The mean-field theory, which serves as the foundation for this work, has been discussed previously~\cite{Ma-Weng2014}, so we will present solely the essential self-consistent equations. Although the number of $b$-spinons is one per site, the number of actual local moments decreases as holes are doped into the system. In the conventional slave-particle theory, the particle number of $b$-spinons is reduced accordingly. However, in previous work~\cite{Ma-Weng2014}, the number of $b$-spinons remains constant, while an auxiliary $a$-spin was introduced to account for the reduction in local moments. A Lagrange multiplier $\gamma_a$ was used to enforce the constraint: $\boldsymbol{S}_{i}^{b} n_{i}^{h} + \boldsymbol{S}_{i}^{a} = 0$, where $\boldsymbol{S}_{i}^{b}$ and $\boldsymbol{S}_{i}^{a}$ represent the spin operators for the $b$-spinons and $a$-spinons, respectively.

\begin{figure}
\centering
\includegraphics[width = 0.95 \linewidth]{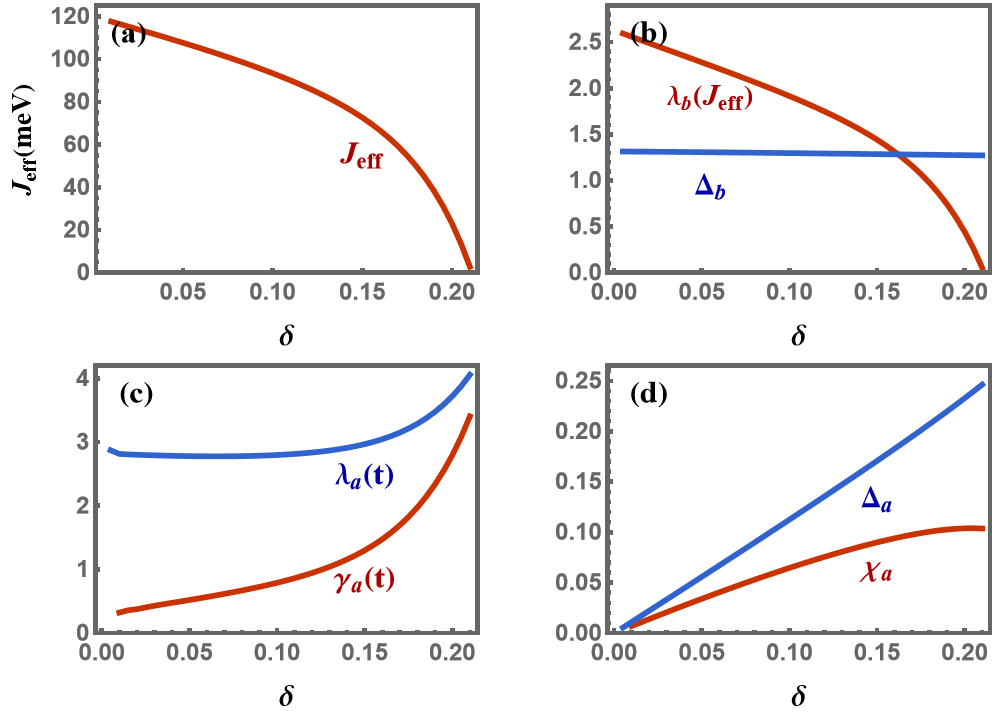}
\caption{
Mean-field order parameters vs. doping density $\delta$. (a) The relation of effective spin interaction $J_{\text{eff}}$ (b) The chemical potential $\lambda_b$ of the $b$-spinons and the pairing strength $\Delta_b$, with the former expressed in units of effective spin interaction 
$J_{\text{eff}}$; (c) The chemical potential $\lambda_a$ of the $a$-spinons and the Lagrange multiplier $\gamma_a$, expressed in units of hopping $t$; (d) $a$-spinon pairing ($\Delta_a$) and hopping ($\chi_a$) amplitudes.
}\label{fig:MFT}
\end{figure}

From the Hamiltonian of Bogolons given in \Eqref{diagB}, the free energy of $b$-spinons is given by:
\begin{equation}
    F_b=\frac{1}{\beta} \sum_{l,k,\sigma,v} \ln 2 \sinh \left[\beta E_{l,k, \sigma, v} / 2\right]+J_{\mathrm{eff}} \Delta_s^2 N-3 \lambda_b N.
\end{equation}
\begin{eqnarray}\label{self2}
&\;&\frac{1}{2}\sum_{l,k,\sigma,v} \frac{\xi_{l,k}^{2} \operatorname{coth}\left(\frac{1}{2} \beta E_{l,k,\sigma,v}\right)}{E_{l,k,\sigma,v}}-2 N (\Delta^{b})^{2} J_{\mathrm{eff}}=0 \notag\\
&\;&\frac{1}{2}\sum_{l,k,\sigma,v} \frac{\lambda_{b} \operatorname{coth}\left(\frac{1}{2} \beta E_{l,k,\sigma,v}\right)}{E_{l,k,\sigma,v}}=3 N
\end{eqnarray}
where $\xi_m$ is determined by the self-consistent equation in Eq.~\eqref{xself}. On the other hand, the free energy for the $a$-spinons is given by:
\begin{align}
F_{a} = & -\frac{2}{\beta} \sum_{k,\alpha=\pm}{}^\prime \ \ln \left[2 \cosh \left(\beta E_{k, \alpha}^a / 2\right)\right] \nonumber \\
& +\gamma_a 2 N\left(\left|\chi^{a}\right|^{2}+\left|\Delta_{a}\right|^{2}\right)+\lambda_{a} N(1-\delta)
\end{align}
where $\sum_k^\prime$ denotes the summation over the reduced Brillouin zone due to the $\pi$-flux folding, and $E_{k,\pm}^a=\sqrt{(\xi_{k,\pm}^a)^2+\Delta_k^2}$ with $\xi_{k,\pm}^a=\pm2 (t_{a}+\gamma_a \chi^a) \sqrt{\cos ^{2} k_{x}+\cos ^{2} k_{y}}+\lambda_{a}$. Next, minimizing this mean-field free energy, i.e., $\partial F_{a} / \partial \chi^{a}=\partial F_{a} / \partial \Delta_{a}=\partial F_{a} / \partial \lambda_{a}=0$, gives rise to the self-consistent equations:
\begin{eqnarray}\label{self1}
\begin{aligned}
\sum_{k, \alpha=\pm}{}^\prime& \gamma_a B_{k, \alpha} A_{k}= N \\
\sum_{k, \alpha=\pm}{}^\prime& (-1)^{\alpha} \sqrt{A_{k}} B_{k, \alpha} \xi_{k, \alpha}^a=2 N \chi^{a}g\\
\sum_{k, \alpha=\pm}{}^\prime& \xi_{k, \alpha}^a B_{k, \alpha}=(1-\delta) N
\end{aligned}
\end{eqnarray}
where $A_{k}=\cos ^{2} k_{x}+\cos ^{2} k_{y}$ and $B_{k, \alpha} = \tanh (\frac{1}{2} \beta E_{k, \alpha}^a) / E_{k, \alpha}^a$. Since the bosonic holons are in the condensation state, their contribution to the free energy may be neglected here such that by minimizing the total free $F_a+F_b$ over $\gamma_a$, one obtains
\begin{equation}\label{self3}
	\delta^{2} |\Delta^{b}|^{2}=|\Delta_{a}|^{2}+4 \chi_{a}^{2}
\end{equation}
The values of the parameters used in the main text, i.e., $\chi_a$, $\Delta_a$, $\lambda_a$, $\Delta^b$, $\lambda_b$, and $\gamma_a$, at different doping concentrations $\delta$, are determined by the self-consistent calculations based on the above equations in Eqs.~\eqref{self1}, \eqref{self2} and \eqref{self3}, under the choice of $t_a= 2J$,
$J = t/3 = 120 \meV$ which is the same as in Ref.~\cite{Ma-Weng2014}. The obtained parameters are shown in \Fref{fig:MFT}.

\section{Electric conductivity in the low-temperature PG regime} \label{sec:PS-conductivity}
In this section, we discuss the derivation of electric conductivity/resistivity in the low-temperature PG phase
where holons have a finite condensation amplitude, and the dominant low-energy excitations are
the holon vortices and $b$-spinons.

After the duality transformation introduced in \Secref{sec:chiral-spinons}, we arrive at the following Lagrangian:
\begin{align} \label{eq:L-b_A^h-A^e}
    L = & \int d^2 r \, \frac{1}{2 \pi^2}\frac{m_h}{\rho_0} [ (\partial_2 A^h_0 - \partial_2 A^h_2)^2 +
    (\partial_0 A^h_1 - \partial_1 A^h_0)^2 ] \nonumber \\
    & -\frac{i}{\pi} \varepsilon_{\mu \nu \lambda} A^h_\mu \partial_\nu A^e_\lambda \nonumber \\
    & + \sum_{i,\sigma,v} b_{i,\sigma,v}^\dagger (\partial_0 - i v A^h_0 + \sigma \mu_B B^e
    +\lambda_b ) b_{i,\sigma,v} \nonumber \\
    & - \frac{J_s}{2} \sum_{i, \alpha, \sigma, v} b_{i+\hat{\alpha},\sigma,v}^\dagger b_{i,-\sigma,-v}^\dagger
    e^{i v A^h_{\alpha}(i)} + h.c.
\end{align}

The mean-field configuration of $A^h_\mu$ (denoted as $\barAh_\mu$) can be determined 
through the variational principle, which gives:
\begin{subequations}
\begin{align}
    & \delta/a^2 = \rho_0 = \frac{1}{\pi} ( \partial_1 \barAh_2 - \partial_2 \barAh_1 ), \\
    & B^e a^2 /\pi = - \sum_{\sigma,v} v \langle b_{i,\sigma,v}^\dagger b_{i,\sigma,v} \rangle_\mathrm{mf}.
\end{align}
\end{subequations}
Here the $\delta$ denotes the number of holons per unit cell and $a$ is the lattice constant.
$\langle \dots \rangle_\mathrm{mf}$ stands for the expectation value from a mean-field $b$-spinon
Hamiltonian:
\begin{align}
    H^b_{\mathrm{mf}} =& \sum_{i,\sigma,v} b_{i,\sigma,v}^\dagger (v \barAh_0 + \sigma \mu_B B^e )
    b_{i,\sigma,v} \nonumber \\
    & -\frac{J_s}{2} \sum_{i,\alpha,\sigma,v} b_{i+\hat{\alpha},\sigma,v}^\dagger b_{i,-\sigma,-v}^\dagger
    e^{i v \barAh_{\alpha}(i)} + h.c.
\end{align}
Note that we have replaced $-i \barAh_0 \rightarrow \barAh_0 > 0$ for $B^e > 0$.
As we are interested in the case with a background magnetic field $B^e$, we shall replace
$A^e_\mu \rightarrow \bar{A}^e_\mu + A^e_\mu$, with $\nabla \times \bar{\bm{A}}^e = B^e$;
we will also expand $A^h$ around its mean-field solution  $A^h_\mu \rightarrow \bar{A}^h_\mu
+ A^h_\mu$.
After integrating out the $b$-spinon in \Eqref{eq:L-b_A^h-A^e}, an effective action of gauge
fields $A^h_\mu$ and $A^e_\mu$ can be obtained, which reads
(to the quadratic order):
\begin{align} \label{eq:S-A^h-A^e}
    & S_\mathrm{eff}[A^h,A^e] \nonumber \\
    & = \frac{1}{\beta \mathcal{V}} \sum_{q=(\omega_n,\bq)} -\frac{i}{\pi}
    A^{h\, T}_{-q}
    \begin{pmatrix}
        0 & -i q_2 & i q_1 \\
        i q_2 & 0 & i \omega_n \\
        -i q_1 & -i \omega_n & 0
    \end{pmatrix}
    A^e_q \nonumber \\
    & + \frac{1}{2}\frac{m_h}{\pi^2 \rho_0} A^{h\, T}_{-q}
    \begin{pmatrix}
        \bq^2 & \omega_n q_1 & \omega_n q_2 \\
        q_1 \omega_n & \omega_n^2 & 0 \\
        q_2 \omega_n & 0 & \omega_n^2
    \end{pmatrix}
    A^h_q
    \nonumber \\
& + A^{h\, T}_{-q}
\begin{pmatrix}
    -\chi^{vv}(q) & i \chi^{v,x}(q) & i \chi^{v,y}(q) \\
    i \chi^{x,v}(q) & K_{xx}(q) & K_{xy}(q) \\
    i \chi^{y,v}(q) & K_{yx}(q) & K_{yy}(q)
\end{pmatrix}
A^h_q.
\end{align}
Here $A^{h/e \, T}_q \equiv (A^{h/e}_0(q), A^{h/e}_1(q), A^{h/e}_2(q))$.
The $b$-spinon correlation functions are defined as:
\begin{subequations}
\begin{align}
    & \chi^{vv}(\tau,\br - \br') = -\langle T V(\tau,\br) V(0,\br') \rangle_\mathrm{mf}, \\
    & \chi^{v,x/y}(\tau,\br - \br') = -\langle T V(\tau,\br) j^{p}_{x/y}(0,\br') \rangle_\mathrm{mf}, \\
    & \chi^{x/y, v}(\tau,\br - \br') = -\langle T j^{p}_{x/y}(\tau,\br) V(0,\br') \rangle_\mathrm{mf}.
\end{align}
\end{subequations}
Here the $b$-spinon vorticity ($A^h$ gauge charge) operator 
\begin{equation}
    V_r = \sum_{\sigma,v} v \, b_{r,\sigma,v}^\dagger b_{r,\sigma,v},
\end{equation}
and the paramagnetic current operator is:
\begin{align} \label{eq:para-j^s}
    j^\p_{\alpha,\br} = & \frac{J_s}{2} \sum_{\sigma,v} i v 
    [ b_{\br+\hat{\alpha},\sigma,v}^\dagger b_{\br,-\sigma,-v}^\dagger 
    e^{i v \barAh_{\br+\hat{\alpha},\br} }  \nonumber \\
    & - b_{\br,-\sigma,-v} b_{\br+\hat{\alpha},\sigma,v} e^{ -i \barAh_{\br+\hat{\alpha},\br} } ].
\end{align}
The $K_{\alpha \beta}$ correlators are defined as:
\begin{subequations}
\begin{align}
K_{\alpha \alpha}(\tau, \br-\br')
=& -\langle T j^\p_{\alpha}(\tau, \br) j^\p_{\alpha}(0,\br') \rangle \nonumber \\
& + \delta(\tau) \delta_{\br,\br'} \langle l_{\alpha,\br} \rangle_{\mathrm{mf}}, \label{eq:K_xx} \\
K_{xy}(\tau, \br - \br')
=& -\langle T j^\p_{x}(\tau,\br) j^\p_{y}(0,\br') \rangle. \label{eq:K_xy}
\end{align}
\end{subequations}
Here
\begin{align}
    l_{\alpha,\br} \equiv & \frac{J_s}{2} \sum_{\sigma,v}
    [ b_{\br+\hat{\alpha},\sigma,v}^\dagger b_{\br,-\sigma,-v}^\dagger 
    e^{i v \barAh_{\br+\hat{\alpha},\br} }  \nonumber \\
    & + b_{\br,-\sigma,-v} b_{\br+\hat{\alpha},\sigma,v} e^{ -i \barAh_{\br+\hat{\alpha},\br} } ]
\end{align}
is involved in the diamagnetic current of $b$-spinon:
$j^\dm_{\alpha,\br} = -l_{\alpha,\br} A^h_{\br+\hat{\alpha},\br}$, and its mean-field expectation value
$\langle l_{\alpha,\br} \rangle_{\mathrm{mf}} = \mathrm{const}$.
Moreover, the $K_{\alpha \beta}$ correlators are related to $b$-spinon conductivity (with respect
to $A^h$) through:
\begin{align}
    \sigma^s_{\alpha \beta}(\omega,\bq)
    = \frac{i}{\omega} K_{\alpha \beta}(\omega+i0_+,\bq).
\end{align}

In order to obtain the electric conductivity, one can first integrate out the $A^h$ field
and obtain an effective action of $A^e$:
\begin{equation}
S_\mathrm{eff}[A^e] = \frac{1}{2} \frac{1}{\beta \mathcal{V}} \sum_{q} A^{e\, T}_{-q} \Pi_{\mu \nu}(q) A^e_q.    
\end{equation}
The electric conductivity is then:
\begin{align}
    \sigma^e_{\mu \nu}(\omega,\bq) = \frac{i}{\omega} \Pi_{\mu \nu}(\omega+i0_+, \bq).
\end{align}
The easiest way to obtain $\Pi_{\mu \nu}$ is by taking a temporal gauge in \Eqref{eq:S-A^h-A^e}:
$A^h_0 = A^e_0 = 0$. After some algebra, one obtains:
\begin{align}\label{rhoe}
    \rho^e = (\sigma^e)^{-1} = \rho^h + \pi^2 
    \begin{pmatrix}
        \sigma^s_{yy} & -\sigma^s_{yx} \\
        - \sigma^s_{xy} & \sigma^s_{xx}
    \end{pmatrix}.
\end{align}
Here $\rho^h(\omega,\bq) = \frac{\omega}{i} \frac{m_h}{\rho_0}$ as holons are condensed. 
In the DC limit, $\rho^h \rightarrow 0$, and 
$\sigma^{s}_{xx} = \sigma^s_{yy}$, $\sigma^s_{xy} = -\sigma^s_{yx}$ due to the $4$-fold rotation
the symmetry of the square lattice, the electric resistivity therefore reads:
\begin{equation}
    \rho^e = \pi^2 \sigma^s.
\end{equation}

\section{Spinon mean-field spectra and its conductivity} \label{sec:b-spinon}
According to Eq.~(8) in the main text, with doping $\delta$, $b$-spinons experience
a $\delta \pi$ $A^h$ flux per plaquette of the square lattice, therefore its unit cell includes
$N_L$ plaquettes. 
After introducing the $k$-states for each sublattice $l$:
\begin{equation}
b_{l,k,\sigma,v} = \frac{1}{\sqrt{N_c}} \sum_{r \in l} e^{-i \bk \cdot \br} b_{r,\sigma,v},
\end{equation}
The $H_b$ can be recast into:
\begin{widetext}
\begin{align}
H_b = \sum_{k,\sigma}
\begin{pmatrix}
B_{k,\sigma,1}^\dagger, & B_{-k,-\sigma,-1}^\dagger
\end{pmatrix}
\begin{pmatrix}
    \lambda_b + v \bar{A}^h_0 + \mu_B B^e & \Delta^b_k \\
    \Delta^b_k & \lambda_b - v \bar{A}^h_0 - \mu_B B^e
\end{pmatrix}
\begin{pmatrix}
    B_{k,\sigma,1} \\
    B_{-k,-\sigma,-1}
\end{pmatrix},
\end{align}
\end{widetext}
with $B_{k,\sigma,v}^\dagger \equiv (b_{1,k,\sigma,1}^\dagger, \dots, b_{N_L,k,\sigma,1}^\dagger)$.
The pairing function $\Delta^b_k$ is hermitian, whose eigenvector is denoted as
$\psi_{l,k}$:
\begin{equation}\label{xself}
    \Delta^b_k \psi_{l,k} = \xi_{l,k} \psi_{l,k}.
\end{equation}
Here the $\xi_{l,k}$'s are essentially flat bands, i.e., LLs.
One can introduce the ``band'' basis through:
\begin{subequations}
\begin{align}
    B_{k,\sigma,1} = & \sum_l \psi_{l,k} \tilde{b}_{l,k,\sigma,1}, \\
    B_{-k,\sigma,-1}^\dagger = & \sum_l \psi_{l,k} \tilde{b}_{l,-k,\sigma,-1}^\dagger.
\end{align}
\end{subequations}
$H_b$ is then diagonalized in the band basis with an intra-band pairing:
\begin{widetext}
\begin{align}
H_b = \sum_l \sum_{k,\sigma} 
\begin{pmatrix}
\tilde{b}_{l,k,\sigma,1}^\dagger, & \tilde{b}_{l,-k,-\sigma,-1}
\end{pmatrix}
\begin{pmatrix}
\lambda_b + v \bar{A}^h_0 + \sigma \mu_B B^e & \xi_{l,k} \\
\xi_{l,k} & \lambda_b - v \bar{A}^h_0 - \sigma \mu_B B^e
\end{pmatrix}
\begin{pmatrix}
\tilde{b}_{l,k,\sigma,1} \\
\tilde{b}_{l,-k,-\sigma,-1}^\dagger
\end{pmatrix}
\end{align}
\end{widetext}
After a Bogoliubov transformation:
\begin{align}
\begin{pmatrix}
\tilde{b}_{l,k,\sigma,1} \\
\tilde{b}_{l,-k,-\sigma,-1}^\dagger
\end{pmatrix}
=
\begin{pmatrix}
    u_{l,k} & -v_{l,k} \\
    v_{l,k} & u_{l,k}
\end{pmatrix}
\begin{pmatrix}
\beta_{l,k,\sigma,1} \\
\beta_{l,-k,-\sigma,-1}^\dagger
\end{pmatrix},
\end{align}
$H_b$ is diagonalized by the Bogolons:
\begin{equation}\label{diagB}
H_b = \sum_l \sum_{k,\sigma} \beta_{l,k,\sigma,v}^\dagger \beta_{l,k,\sigma,v} E_{l,k,\sigma,v}
+ \mathrm{const}.
\end{equation}
Here 
\begin{subequations}
\begin{align}\label{Emb}
E_{l,k} = & \sqrt{\lambda_b^2 - \xi_{l,k}^2} \\
E_{l,k,\sigma,v} = & E_{l,k} + v \bar{A}^h_0 + \sigma \mu_B B^e.
\end{align}
\end{subequations}
\begin{subequations}
\begin{align}
u_{l,k} = & \frac{1}{\sqrt{2}} \sqrt{\frac{\lambda_b}{E_{l,k}} + 1}, \\
v_{l,k} = & \sgn(\xi_{l,k}) \frac{1}{\sqrt{2}} \sqrt{\frac{\lambda_b}{E_{l,k}} - 1}.
\end{align}
\end{subequations}
Plots of $E_{l,k,\sigma,v}$ at $\delta = 0.125$ is shown in Fig.2(c) of the main text.

The vorticity Hall conductivity $\sigma^s_{xy}$ involves the correlation function about
paramagnetic current operator $j^\p_{\alpha,r}$, according to \Eqref{eq:para-j^s}, the DC ($\bq = 0$)
current operator reads:
\begin{align}
& j^p_{\alpha,\bq = 0} = \sum_{k} \sum_{m,n} 
\psi_{m,k}^\dagger \frac{\partial \Delta^b_k}{\partial k_\alpha} \psi_{n,k} \nonumber \\
& \begin{pmatrix}
    \tilde{b}_{m,k,\sigma,1}^\dagger, & \tilde{b}_{m,-k,-\sigma,-1}
\end{pmatrix}
\tau^x
\begin{pmatrix}
    \tilde{b}_{n,k,\sigma,1} \\
    \tilde{b}_{n,-k,-\sigma,-1}^\dagger
\end{pmatrix}.
\end{align}
Here $\tau^x$ is the Pauli-$X$ matrix. Accodring to \Eqref{eq:K_xy},
\begin{widetext}
\begin{align}
& K_{xy}(i\nu_n, \bq = 0) \nonumber \\
& = \frac{1}{N} \sum_{k,\sigma} \sum_{m,n}
\psi_{n,k}^\dagger \frac{\partial \Delta^b_k}{\partial k_x} \psi_{m,k} \,
\psi_{m,k}^\dagger \frac{\partial \Delta^b_k}{\partial k_y} \psi_{n,k} \times \nonumber \\
& \left[ (u_{n,k} v_{m,k} + v_{n,k} u_{m,k} )^2 
\left( \frac{ n_B(E_{n,k,\sigma,1}) - n_B(E_{m,k,\sigma,1}) }{i \nu_n + E_{n,k}-E_{m,k}}
+ \frac{n_B(E_{m,k,-\sigma,-1}) - n_B(E_{n,k,-\sigma,-1})}{i \nu_n + E_{m,k}-E_{n,k}} \right) \right. \nonumber \\
& \left. + (u_{n,k} u_{m,k} + v_{n,k} v_{m,k} )^2 \left(
\frac{1+n_B(E_{m,k,\sigma,1}) + n_B(E_{n,k,-\sigma,-1})}{i\nu_n-E_{m,k}-E_{n,k}}
- \frac{1+n_B(E_{m,k,-\sigma,-1}) + n_B(E_{n,k,\sigma,1})}{i\nu_n + E_{m,k} + E_{n,k}}
\right) \right].
\end{align}
\end{widetext}
After some algebra, it can be shown that:
\begin{align}
\sigma^s_{xy} =& \lim_{\omega \rightarrow 0} \frac{i}{\omega+i0_+} K_{xy}(\omega+i0_+,\bq = 0) \nonumber \\
=& \sum_{n,\sigma,v} \int \frac{d^2 k}{(2\pi)^2} \mathcal{F}^n_{xy,k} \,v \, n_B(E_{n,k,\sigma,v})
\end{align}

\subsection{Longitudinal conductivity}
The $b$-spinons' longitudinal vorticity conductivity $\sigma^s$ reads:
\begin{widetext}
\begin{align} \label{eq:sigma^s_xx}
    \sigma^s_{\alpha,\alpha} = & \frac{\pi}{N} \sum_{k} \sum_{\sigma,v} \sum_{l,m} 
    | \psi_{l,k}^\dagger \frac{\partial \Delta^b_k}{\partial k_\alpha} \psi_{m,k}|^2 \times
    \beta \left \{ n_B(E_{l,k,\sigma,v}) [ 1+n_B(E_{l,k,\sigma,v}) ] A(E_{l,k}-E_{m,k}) ( u_{l,k} v_{m,k} + v_{l,k} u_{m,k}  )^2 \right. \nonumber \\
    & \left. - n_B(E_{l,k,\sigma,v}) [ 1+n_B(E_{l,k,\sigma,v}) ] A(E_{l,k} + E_{m,k}) ( u_{l,k} u_{m,k} + v_{l,k} v_{m,k}  )^2 \right \}.
\end{align}
\end{widetext}
Here the Lorentzian function $A(x) = (\Gamma/\pi)/(x^2 + \Gamma^2)$ is introduced through a broadened $b$-Bogolon spectral function
(to account for scattering effect)~\cite{Ma-Weng2014},
the broadening factor is set to be $\Gamma = 0.02 J = 2.4 \meV$ in this study.
At very low temperatures, only the $E_{1,k,\downarrow,-1} \approx E_{1,\downarrow,-1}$
band has significant occupation (see illustration of energy levels in Fig.~2(c)) and contributes most in \Eqref{eq:sigma^s_xx}.
As $n_B(E_{1,k,\downarrow,-1})$ is also fixed by external magnetic field through the constraint on vorticity (Eq.~(9) in the main text),
it follows that as $T \rightarrow 0$, $\sigma^s_{xx}$ is equal to $\beta = 1/(k_B T)$ times a constant factor determined by $B^e$.
This explains the $1/T$-divergent behavior of $\rho^e$ at low temperatures.
Note that other ways of incorporating the scattering effect, e.g., $\Gamma$ depends on $T$ or vortex corrections,
could render $\rho^e$ finite as $T \rightarrow 0$.

\subsection{Hall angle in the low-temperature PG phase}
\begin{figure}
\centering
\includegraphics[width = 0.95 \linewidth]{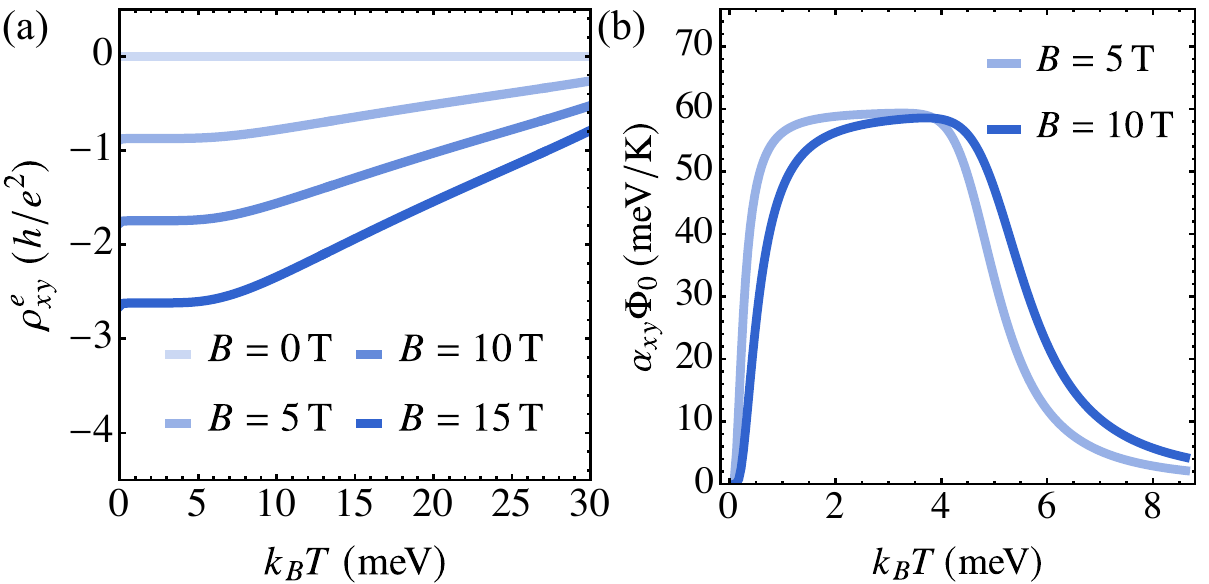}
\caption{
(a) $\rho^e_{xy}$ at $\delta = 0.2$ with different magnetic fields.
At finite $B$, $\rho^e_{xy}$ saturates to a constant at low temperatures.
(b) Nernst data at $\delta = 0.18$.
}\label{fig:rho_xy}
\end{figure}

\begin{figure}
\centering
\includegraphics[width = 0.7 \linewidth]{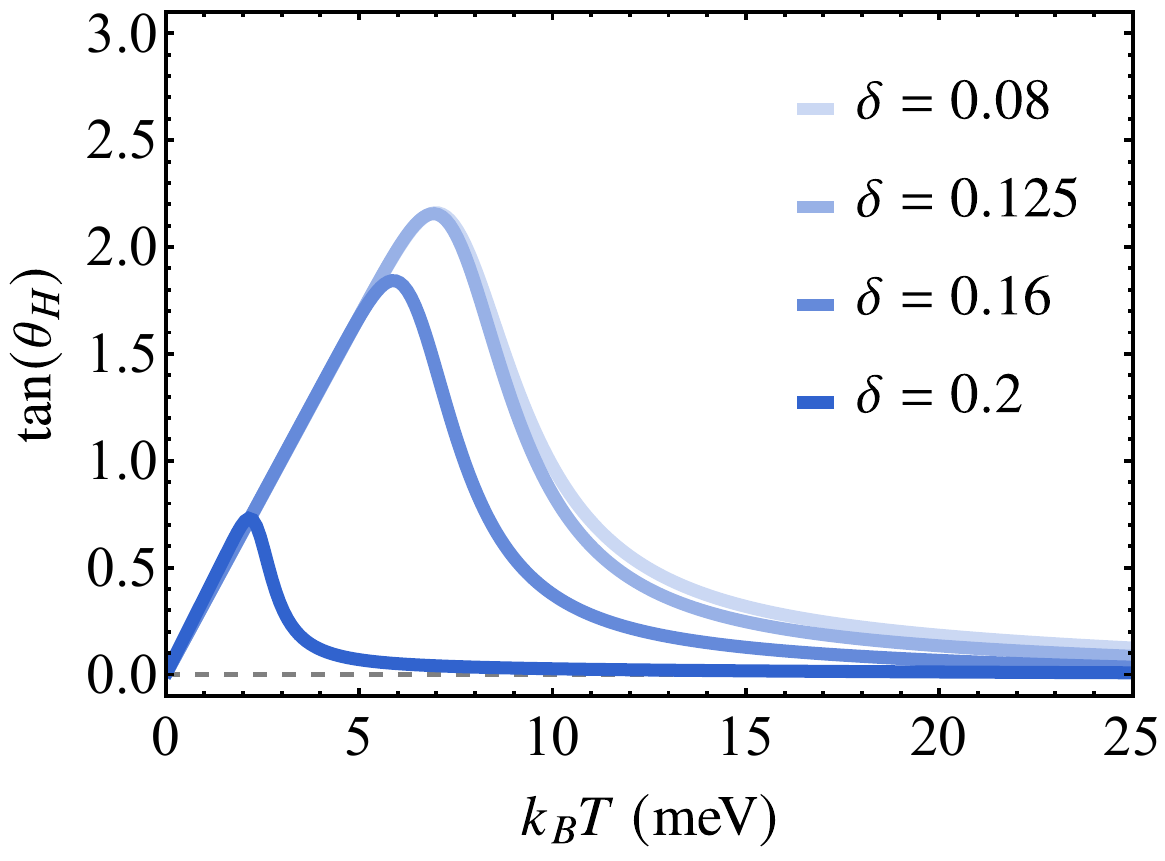}
\caption{
Hall angle in the low-$T$ PG phase at different dopings.
Here the calculation is done at $B^e = 10 \, \mathrm{T}$.
}\label{fig:theta_H}
\end{figure}

The Hall angle in the low-$T$ PG regime at various doping levels is shown in \Fref{fig:theta_H}.
As $T \rightarrow 0$, $\rho^e_{yx}$ increases monotonically (see \Fref{fig:rho_xy}(a)),
while $\rho^e_{xx}$ initially decreases before rising again (see Figs.~3 of the main text).
As a result, $\tan(\theta_H)$ first increases and then gets suppressed.
Considering a more realistic scattering mechanism for the $b$-spinons (so that $\rho^e_{xx}(T=0) \neq 0$) could lead to
saturation of $\tan( \theta_H )$ at finite values.

\section{Linear-$T$ resistivity from $\rho^h$ in the SM phase} \label{sec:rho^h}
In the SM regime, $b$-spinons are no-longer RVB paired and behaves as free local magnetic moments, thereby
producing randomized $A^s$ ``magnetic'' flux felt by holons.
The holons' Hamiltonian reads:
\begin{equation}
    H_h = \sum_{r} \lambda^h h_r^\dagger h_r - t_h \sum_{r,\alpha} h_{r+\hat{\alpha}}^\dagger h_{r} e^{ i [ A^s_{\alpha}(r) + A^e_{\alpha}(r) ] } + h.c.
\end{equation}
The current operator $j_\alpha = j^\p_\alpha + j^\dm_\alpha$ contains the paramagnetic and diamagnetic parts, which are defined as:
\begin{subequations}
\begin{align}
j^\p_\alpha(r) = & i t_h \left( h_{r+\hat{\alpha}}^\dagger h_r e^{i A^s_{\alpha}(r) } - h_{r}^\dagger h_{r+\hat{\alpha}} e^{-i A^s_{\alpha}(r)} \right),
\label{eq:j^hp} \\
j^\dm_\alpha(r) = & -T_\alpha(r) A^e_\alpha(r), \label{eq:j^hd} \\
T_\alpha(r) \equiv & t_h \left( h_{r+\hat{\alpha}}^\dagger h_{r} e^{i A^s_\alpha(r)} + h_{r}^\dagger h_{r+\hat{\alpha}} e^{-i A^s_{\alpha}(r) } \right).
\end{align}
\end{subequations}
The electrical conductivity (at a given $A^s$ configuration) is defined as:
\begin{equation} \label{eq:sigma^h-def}
\sigma^e_{\alpha \alpha}(\omega, \bm{q}) = \frac{i}{\omega} \left[ \mathcal{K}_{\alpha \alpha}(\omega+i 0_+,\bm{q}) + \langle T_\alpha(r) \rangle_0 \right].
\end{equation}
The paramagnetic current-current correlation function is defined as:
\begin{align}
\mathcal{K}_{\alpha \alpha}(\tau, \bm{q}) = & -\frac{1}{N} \langle T j^\p_{\alpha}(\tau, \bm{q}) j^\p_{\alpha}(0, -\bm{q}) \rangle_0, \nonumber \\
\mathcal{K}_{\alpha \alpha}(i \nu_n, \bm{q}) = & \int_{0}^\beta d \tau \, e^{i \nu_n \tau} \mathcal{K}_{\alpha \alpha}(\tau,\bm{q}).
\end{align}
Here $N$ is the number of lattice sites.
The average $\langle \dots \rangle_0$ is taken with respect to the Hamiltonian without $A^e$: $H_h^0 \equiv H_h(\bm{A}^e = 0)$,
which can be diagonalized by the single-particle eigenmodes $d_j$:
\begin{align}
\begin{pmatrix}
    h_1 \\
    \vdots \\
    h_N
\end{pmatrix}
 & = 
 \begin{pmatrix}
     w_1 & w_2 & \dots & w_N
 \end{pmatrix}
 \begin{pmatrix}
     d_1 \\
     \vdots \\
     d_N
 \end{pmatrix}
,
\end{align}
and the holon Hamiltonian can be written as:
\begin{equation}
    H_h^0 = \sum_j d_j^\dagger d_j \epsilon_j.
\end{equation}
According to the definition in \Eqref{eq:j^hp}, the paramagnetic current operator
\begin{align}
j^p_\alpha(\bm{q}=0) = & \sum_{r} j^p_\alpha(r) \nonumber \\
 = & 
\begin{pmatrix}
 h_1^\dagger & \dots & h_N^\dagger
\end{pmatrix}
M^\alpha
\begin{pmatrix}
 h_1 \\
 \vdots \\
 h_N
\end{pmatrix}
\nonumber \\
 = & \sum_{m,n} w_m^\dagger M^\alpha w_n d_m^\dagger d_n.
\end{align}
Here the matrix $M^\alpha$ is defined as:
\begin{equation}
M^\alpha_{r,r'} = 
\begin{cases}
    \ i t_h e^{i A^s_\alpha(r')}, & r = r'+\hat{\alpha} \\
    \ -i t_h e^{-i A^s_\alpha(r)}, & r' = r + \hat{\alpha} \\
    \ 0, & \mathrm{others}
\end{cases}
\end{equation}
It can be shown that,
\begin{align}
\mathcal{K}_{\alpha \alpha}(i \nu_n, 0)
= & \frac{1}{N} \sum_{m,n} w_m^\dagger M^\alpha w_n w_n^\dagger M^\alpha w_m \nonumber \\
& \times \frac{n_B(\epsilon_m) - n_B(\epsilon_n)}{i \nu_n + \epsilon_m - \epsilon_n}.
\end{align}
From \Eqref{eq:sigma^h-def}, the real part of the conductivity $\sigma^{e, \mathrm{I}}_{\alpha \alpha}$ reads:
\begin{align}
\sigma^{e,\mathrm{I}}_{\alpha \alpha}(\omega) = & \frac{1}{N} \sum_{m,n} w_m^\dagger M^\alpha w_n w_n^\dagger M^\alpha w_m \nonumber \\
& \times \frac{n_B(\epsilon_m) - n_B(\epsilon_m+\omega)}{\omega} \pi \delta(\omega+\epsilon_m - \epsilon_n),
\end{align}
Taking the $\omega \rightarrow 0$ limit, one obtains the DC conductivity:
\begin{align}
\sigma^{e, \mathrm{I}}_{\alpha \alpha} = &
\frac{1}{N} \sum_{m,n} w_m^\dagger M^\alpha w_n w_n^\dagger M^\alpha w_m \nonumber \\
& \times \beta n_B(\epsilon_m) [ 1 + n_B(\epsilon_m) ] \pi \delta(\epsilon_m - \epsilon_n).
\end{align}
Finally, we should average over different $A^s$ flux configurations to get the physical conductivity:
$\langle \sigma^{e,\mathrm{I}}_{\alpha \alpha} \rangle_{\bm{A}^s}$.
A plot of $\rho^e$ at doping $\delta = 0.2$ is shown in Fig.~4(b) of the main text.

\section{Specific heat and spin susceptibility of $b$-spinons} \label{sec:heat}
From the Hamiltonian of Bogolons given in \Eqref{diagB}, the free energy of $b$-spinons is given by:
\begin{align}
    F_b = & \frac{1}{\beta} \sum_{l,k,\sigma,v} \ln 2 \sinh \left[\beta E_{l,k, \sigma, v} / 2\right] \nonumber \\
    & +J_{\mathrm{eff}} \Delta_s^2 N-3 \lambda_b N.
\end{align}
Then, the contribution to the specific heat from $b$-pinons can be expressed as:
\begin{align}
    \gamma &\equiv C_v^b / T =-\frac{1}{N} \frac{\partial^2}{\partial T^2} F_b\\
    &=\frac{1}{N} \sum_{l,k, \sigma, v} \frac{E_{l,k, \sigma, v}^2}{k_B T^3} n_B\left(E_{l,k, \sigma, v}\right)\left[n_B\left(E_{l,k, \sigma, v}\right)+1\right]\notag
\end{align}
here $n_B(\omega)=1 /\left(e^{\beta \omega}-1\right)$ denotes the bosonic distribution function. Similarly, the total magnetic moment induced by the magnetic field from $b$-spinons can be expressed as:
\begin{equation}
    M_b=\mu_B \sum_{l,k,v}\left[n_B\left(E^b_{l,k,\uparrow,v}\right)-n_B\left(E^b_{l,k,\downarrow,v}\right)\right]
\end{equation}
 Therefore, the spin suscepbility $\chi^s$ at local site is defined by
\begin{align}
    \chi^s = & \frac{M_b}{N B} \mid_{B \rightarrow 0} \nonumber \\
    = & \frac{1}{N}\sum_{l,k,\sigma,v}  \mu_B^2 \beta  n_B\left(E_{l,k,\sigma,v}\right)\left[n_B\left(E_{l,k,\sigma,v}\right)+1\right]
\end{align}

\bibliography{reference.bib}




\end{document}